 \documentclass[final,5p,times,twocolumn]{elsarticle}

\usepackage{amsmath,amssymb,amsfonts}
\usepackage{mathrsfs}
\usepackage{chemformula}
\usepackage{color}
\usepackage[normalem]{ulem}
\usepackage{stmaryrd}
\usepackage{diagbox}
\usepackage[parfill]{parskip}
 \biboptions{sort&compress}

\newcommand{\pfr}[2]{\ensuremath{\frac{\partial #1}{\partial #2}}}

\newcommand{\ep}{\epsilon}
\newcommand\Lew{\mbox{\textit{Le}}}
\newcommand\Pra{\mbox{\textit{Pr}}}

\newcommand{\vect}[1]{\mathbf{#1}}

\journal{Proceedings of the Combustion Institute}

\begin{document}

\begin{frontmatter}

\title{Premixed flame quenching distance between cold walls: Effects of flow and Lewis number}

\author{Aiden Kelly$^a$, Remi Daou$^b$, Joel Daou$^{a}$, Vadim N. Kurdyumov$^c$, Prabakaran Rajamanickam$^d$}
\address{$^a$Department of Mathematics, University of Manchester, Manchester M13 9PL, UK \\
$^b$Universit\'e Saint-Joseph de Beyrouth, Facult\'e d'Ing\'enierie et d'Architecture, ESIB, Lebanon\\
$^c$Department of Energy, CIEMAT, Madrid 28040, Spain\\
$^d$Department of Mathematics and Statistics, University of Strathclyde, Glasgow G1 1XQ, UK}

\begin{abstract}
This study investigates the critical conditions for flame propagation in channels with cold walls. We analyse the impact of the Lewis number and flow amplitude ($A$) on the minimum channel width required to sustain a premixed flame. Our results span a wide range of Lewis numbers, encompassing both aiding and opposing flow conditions. Results are presented for both variable and constant density models. A combined numerical approach, involving stationary and time-dependent simulations, is employed to determine quenching distances and solution stability. We find that smaller Lewis numbers and aiding flows ($A < 0$) facilitate flame propagation in narrower channels, while opposing flows ($A > 0$) tend to destabilise the flame, promoting asymmetric solutions. For sufficiently large positive values of 
$A$, the quenching distance is determined by asymmetric solutions, rather than the typical symmetric ones.
\end{abstract}


\end{frontmatter}

\section{Introduction\label{sec:introduction}}

The quenching distance $d$ may be defined as the minimum separation between parallel flat plates containing a combustible mixture in which a flame can still propagate~\cite[p.~269]{williams2018combustion}.
The primary  factor in determining the quenching distance is the conductive heat loss to the plates. Consequently, under ideal adiabatic conditions, a   quenching distance $d\to 0$ can be theoretically achieved.  However, in practice, $d$ is typically several times larger than the laminar adiabatic flame thickness $\delta_L$. The quenching distance is also strongly affected by mixture composition as shown in early works such as \cite{spalding1957theory}. Previous studies on hydrocarbon fuels have reported quenching distances with $d/\delta_L$ ratios ranging from 30 to 50~\cite{mayer1957theory,lewis2012combustion,williams2018combustion,jarosinski1983flame,kim2006numerical}. However, by employing sub-unity Lewis number fuels like hydrogen, this ratio can be significantly reduced. The diffusive-thermal effects in such fuels promote multi-dimensional flame structures, enabling them to withstand substantial heat losses~\cite{fernandez2018analysis,veiga2020unexpected,gu2021propagation}.

 This study will focus on the extreme case where the plates are held at the temperature of the cold gas, as this will provide an upper bound for quenching distances. Although previous studies have considered cold wall conditions experimentally~\cite{jarosinski1983flame}, theoretically~\cite{lewis2012combustion,von1953thermal}, and numerically~\cite{daou2002influence,kim2006numerical, tsai2008asymmetric},  
 dedicated studies systematically investigating quenching distance in terms of Lewis number and flow strength are yet to be conducted. This paper aims to contribute to bridging this gap by determining quenching distances across a range of Lewis numbers and flow intensities.   Particular attention will be paid to the existence of various symmetric and asymmetric solutions and their stability. The investigation should therefore complement recent findings in the literature concerned with flame propagation and stability in channels under varying flow conditions and different thermal wall properties~\cite{kelly2024threedimensional,kurdyumov2011lewis,kurdyumov2016structure,kurdyumov2000flame,daou2001flame,ronney2003analysis,ronney1998understanding}.

\section{Governing equations\label{sec:equations}} 
\begin{figure}[h!]%make text bigger
\centering
\includegraphics[width=192pt]{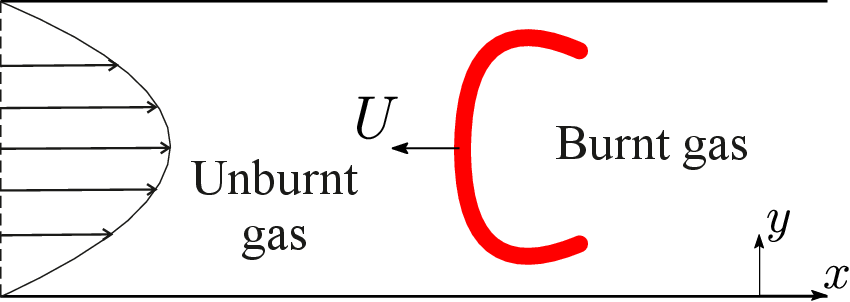}
\caption{\footnotesize A schematic representation of a premixed flame propagating through a Poiseuille flow at (non-dimesnional) speed $U$ relative to the channel walls, which are maintained at the unburnt gas temperature. 
}
\label{fig:setup}
\end{figure}
Consider a reactive mixture   between two parallel plates held at  the unburnt gas temperature $T_u$, in the presence of  a Poiseuille flow given far upstream by
\begin{equation*}
    \vect u^* = A^*\left(1-\frac{y^{*^2}}{h^2}\right) \vect e_{x}  \label{eq:velground}
\end{equation*}
as depicted in Fig.~\ref{fig:setup}. Here, $A^*$ denotes the dimensional flow amplitude 
and $h$ the channel half-width.  A one-step chemistry model with reaction rate $\omega^*= \rho B Y_F e^{-E/RT}$ is adopted  where $B$ is the pre-exponential factor, $\rho$ the gas density, $Y_F$  the fuel mass fraction assumed deficient, $E$   the activation energy, $R$ the universal gas constant, and $T$   the gas temperature.    Thermal conductivity $\lambda$, heat capacity $c_p$, dynamic viscosity $\mu$, and the product of the density with  the fuel diffusion coefficient $D_F$ are assumed  constant.   As reference quantities for nondimensionalisation we select the channel half-width $h$, the laminar flame speed $S_L$ (whose thickness is denoted by $\delta_L$), and the density $\rho_u$ (where the subscript $u$ denotes throughout conditions in the unburnt gas). With the preceding assumptions, the problem is governed by the  non-dimensional equations 
\begin{align*}
    \pfr{\rho}{t}+\nabla\cdot(\rho\vect u)&=0 \\
    \rho \pfr{\vect u}{t}+ \rho \vect u\cdot \nabla\vect u &= - \nabla p + \frac{\Pra}{\ep} \nabla^2\vect{u} \\
    \rho \pfr{\theta}{t}+ \rho \vect u\cdot \nabla\theta &= \frac{1}{\ep}\nabla^2\theta + \ep\,\omega  \\
     \rho \pfr{y_F}{t}+ \rho \vect u\cdot \nabla y_F &= \frac{1}{\ep\Lew}\nabla^2 Y - \ep\,\omega  
\end{align*} 
Here $p$ is  the (modified) pressure (scaled with $\rho_u S_L^2$), $\operatorname{Pr}=\mu c_{p} / \lambda$  the Prandtl number and $\theta=\left(T-T_{u}\right) /\left(T_{\mathrm{ad}}-T_{u}\right)$ the nondimensional temperature;   $T_{\text {ad }}$ is the   adiabatic flame temperature; $\rho$ and $\theta$ are linked by the ideal-gas equation of state 
\begin{equation*}
\rho=\left(1+\frac{\alpha}{1-\alpha} \theta\right)^{-1}  
\end{equation*}
where $\alpha=(T_{ad}-T_u)/T_{ad}$. The non-dimensional reaction rate is given by
\begin{equation*}
    \omega = \frac{\beta^2}{2\Lew(1-\alpha)}\rho \, y_F\exp\left[\frac{\beta(\theta-1)}{1+\alpha(\theta-1)}\right] 
\end{equation*}
with the Zeldovich number $\beta=E(T_{ad}-T_u)/RT_{ad}^2$. For $S_L$, the large-$\beta$ asymptotic expression   
\[ 
   S_L^2 = 2\Lew\beta^{-2} B D_{T,u} (1-\alpha)e^{-E/RT_{ad}} \]
is adopted with the Lewis number being defined  by $\Lew=D_T/D_F = \lambda/\rho c_p D_F$. The laminar flame thickness is $\delta_L=D_{T,u}/S_L$ and is used to define the scaled half-channel width $\epsilon=h/\delta_L$.

The boundary conditions in the far field are 
\begin{align*}
    x \to -\infty&: \quad {\bf u} = A (1-y^2) {\bf e}_x \quad \theta=0 \quad y_F=1 \\
    x \to +\infty&: \quad \pfr{\theta}{x}=\pfr{y_F}{x}=0 
\end{align*}
where $A=A^*/S_L$ is the scaled flow amplitude. At the plates,  no-slip conditions are imposed along with the impermeable cold walls condition 
\begin{equation*}
   \quad\pfr{y_F}{y} =0 \quad \text{and} \quad \theta = 0  \quad {\rm at} \quad   y=\pm1
\end{equation*}

The problem is numerically investigated by identifying stationary solutions and determining  their non-dimensional propagation speed $U$ (relative to the channel wall) or, more significantly, their effective propagation  speed, $\widetilde{U}\equiv U+2A/3$, relative to the unburnt gas mean flow. The stability of these solutions is then assessed through time-dependent simulations.

\section{Results\label{sec:Results}}

The computations are carried out using the finite-element package COMSOL Multiphysics. The domain non-dimensional size is 2 in the $y$-direction and ranges from 50 to 300 in the $x$-direction. Grids of typically 500000  triangular elements are used, including local refinement around the reaction zone. A technical point to note is that $U$, which appears as an unknown parameter (or eigenvalue), is determined so that the flame remains anchored to the origin of the computational domain. This is achieved by imposing the constraint $\theta=0.5$ at the origin, as done in~\cite{daou2023flame,daou2023premixed}. For the simulations, both stationary and time-dependent, we adopt the fixed values $\alpha=0.85$, $\Pra=1$ and $\beta=10$ and vary the parameters $\Lew$, $\ep$ and $A$.

\subsection{Steady solutions in the absence of flow, $A=0$\label{sec:ResultsA0}} 

\begin{figure}[h!]
\centering
\includegraphics[width=192pt]{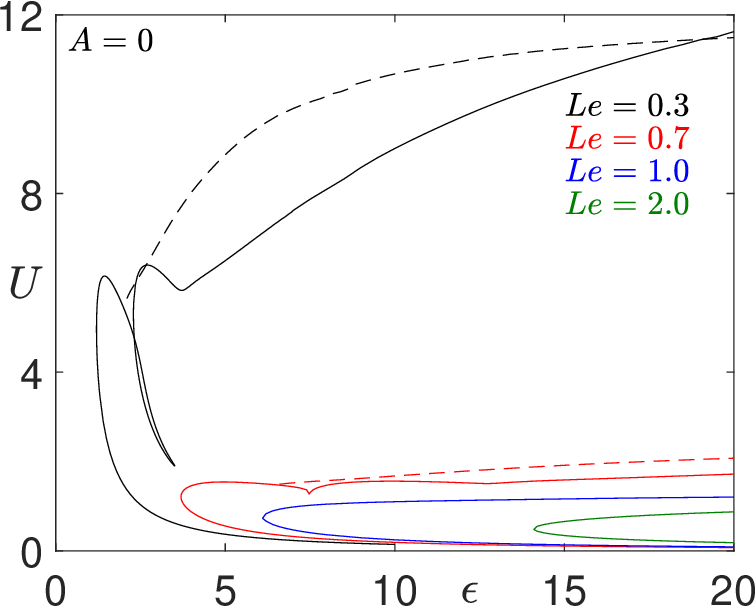}
\caption{\footnotesize The propagation speed $U$ versus $\epsilon$ for $A=0$ and selected values of  $\Lew$. Solid lines represent solutions symmetric with respect to $y=0$ and dashed lines asymmetric solutions.}
\label{fig:VDA0CCurves}
\end{figure}

\begin{figure*}[h!]
\centering
\includegraphics[width=408pt]{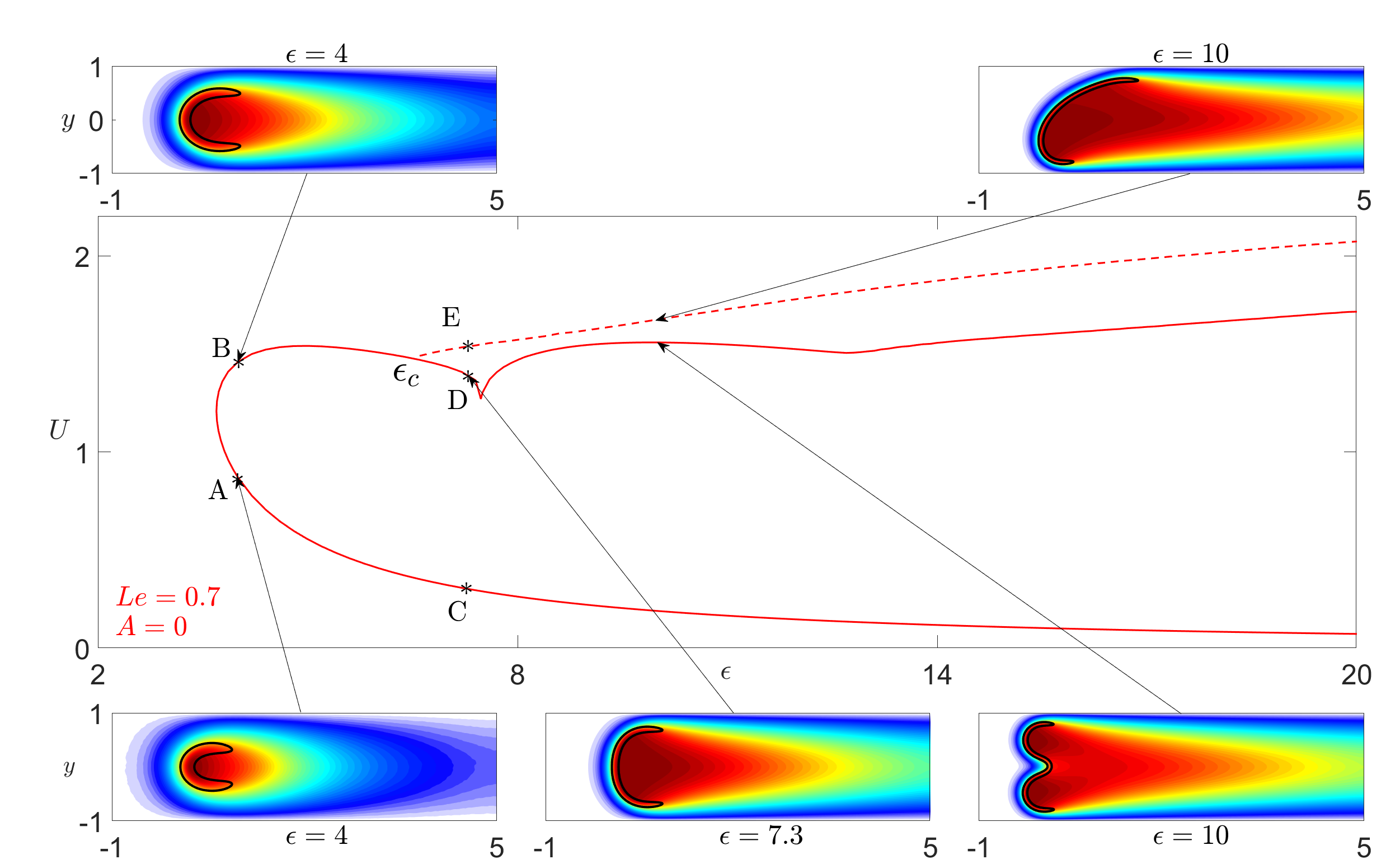}
\caption{\footnotesize The propagation speed $U$ versus $\epsilon$ for $A=0$ and $\Lew=0.7$. As in the previous figure, solid lines represent solutions symmetric with respect to $y=0$ and dashed lines represent asymmetric solutions. These symmetric and asymmetric solutions are illustrated by the insets for selected values of $\epsilon$. Shown in each inset are colour-coded temperature fields as well as a single reaction rate contour ($\omega=0.1 \,\omega_{\max}$)  marked by a black line to locate the reaction zone. The points marked by asterisks labelled by capital letters A to E refer to solutions which will be discussed later in section~\ref{sec:ResultsStab} on stability. }
\label{fig:VDLe07A0epsUDiag}
\end{figure*}

We begin with the case $A=0$, whose results are presented in Fig.~\ref{fig:VDA0CCurves}  and Fig.~\ref{fig:VDLe07A0epsUDiag}. Shown  are curves depicting $U$ versus $\epsilon$ for selected values of $\Lew$ corresponding to stationary solutions of the governing equations. The solid curves correspond to solutions found to be symmetric with respect to the $y=0$ plane. The dashed curves represent asymmetric solutions, and are found only for the sub-unit Lewis number cases when $A=0$. Such symmetric and asymmetric solutions are illustrated in the colour insets of Fig.~\ref{fig:VDLe07A0epsUDiag} pertaining to $\Lew=0.7$, where temperature fields and reaction rate contours are plotted for selected values of $\epsilon$. It is seen for this case, that asymmetric solutions exist  for values of $\epsilon$ larger than a critical value, $\epsilon_c \approx 6.5$, associated with the bifurcation point corresponding to the intersection of the solid and dashed lines. For $\epsilon<\epsilon_c$, two solutions are depicted (for $\epsilon=4$) with both being single-headed symmetric flames. The stronger  burning solution is located on the upper branch and corresponds to a  larger flame.  As $\epsilon$ increases beyond $\epsilon_c$, the   symmetric flame on the upper solid branch  splits into a double-headed flame  at a value of $\epsilon \approx 7.47$, corresponding to the cusp point in the figure.  As the figure indicates, for $\epsilon>\epsilon_c$, asymmetric flames also exist as illustrated in the inset corresponding to $\epsilon=10$. It is worth noting that in general the solid curves lower branches represent slower-propagating flames and that these are found to be unstable by time-dependent simulations, as will be demonstrated in section~\ref{sec:ResultsStab}. Their upper branches are found typically to represent stable (symmetric) solutions for $\epsilon <\epsilon_c$ where no asymmetric solutions exist, and unstable otherwise. The asymmetric solution on the other hand are found to be stable. For a given value of the Lewis number and the flow amplitude $A$ (with $A=0$ in this section), the smallest value of $\epsilon$ on the associated solution curve, which marks a turning point defines the quenching distance.

\begin{figure}[h!]
\centering
\includegraphics[width=192pt]{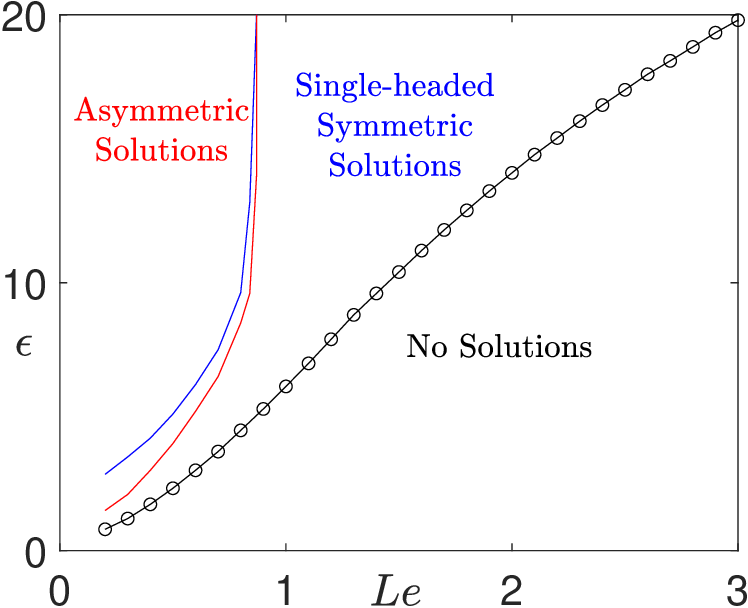}
\caption{\footnotesize Existence domains of the steady solutions for $A=0$. Quenching distance: lower black curve. Asymmetric solutions: left of red curve. Symmetric solutions: left of black curve; these symmetric solutions are single-headed flames to the right of the blue curve and multi-headed flames to its left.  
}
\label{fig:VDA0LeCriteps}
\end{figure}

A synthetic presentation of the results, focusing on the types of solutions and their existence domains in  the $\Lew$-$\epsilon$ plane,  is given in Fig.~\ref{fig:VDA0LeCriteps}. The lower black curve in this figure determines  the quenching distance. Asymmetric solutions are located to the left of the red curve.  Symmetric solutions are found to the left of the black curve; these symmetric solution are single-headed in the region bounded by the blue and black curves and multi-headed to the left of the blue curve. Therefore, between the blue and red curves both asymmetric and single-headed symmetric solutions exist.  We note that all curves, including the quenching curve, increase with the Lewis number, with the latter showing nearly linear behaviour. Notably, asymmetric solutions do not exist for $\Lew \gtrsim 0.85$ (when $A=0$).
 
%\newpage 
\subsection{Effect of the flow amplitude $A$ on the steady solutions\label{sec:ResultsAn0}} 

\begin{figure}[h!]
\centering
 \includegraphics[width=192pt]{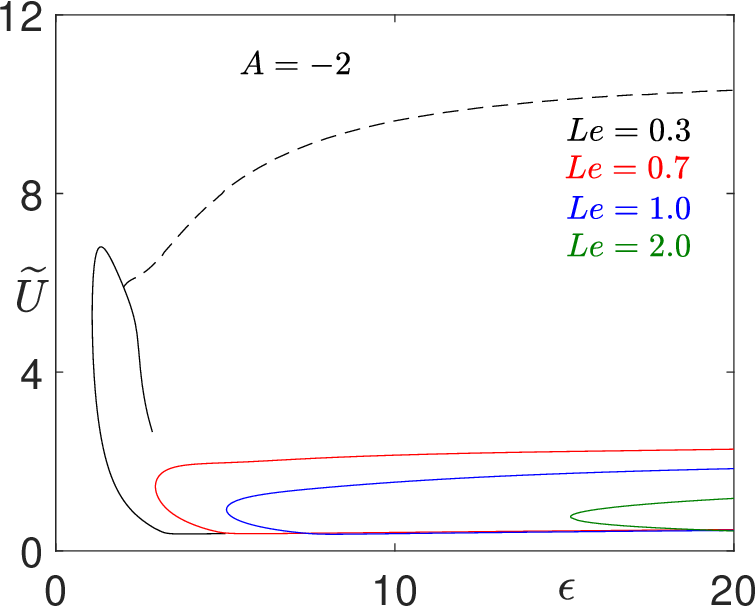}
\includegraphics[width=192pt]{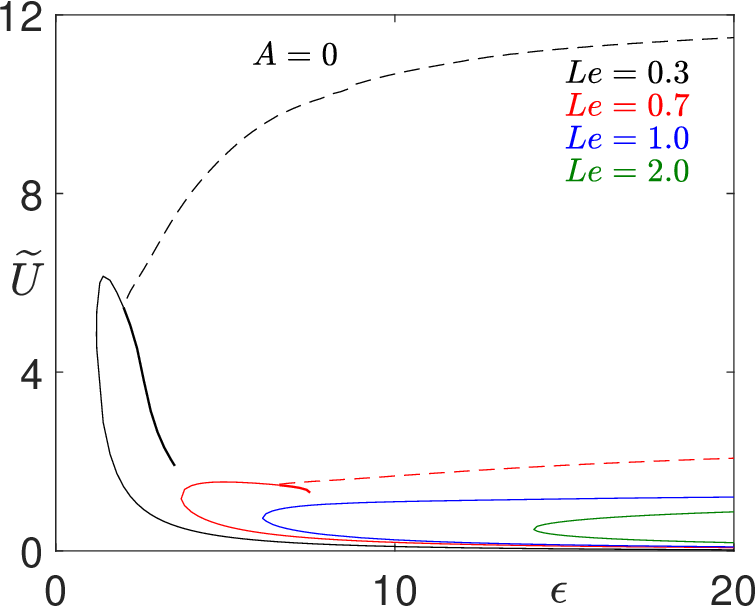}
 \includegraphics[width=192pt]{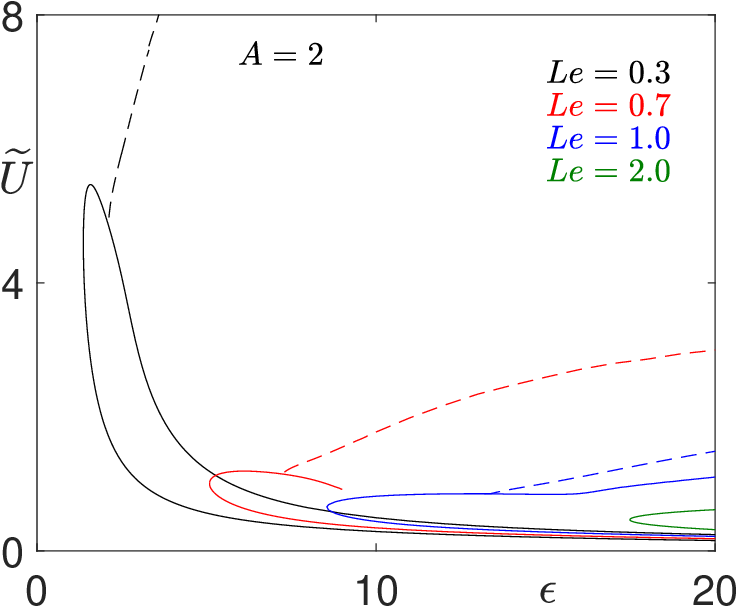}
\caption{\footnotesize The effective propagation speed $\widetilde{U}\equiv U+2A/3$ versus $\epsilon$ for  selected values of  $\Lew$. The top, middle and lower subfigures correspond to $A=-2$, $A=0$ and $A=2$, respectively. Solid lines represent solutions symmetric with respect to $y=0$ and dashed lines asymmetric solutions.}
\label{fig:VDAn0CCurves}
\end{figure}

In this section, we examine  the influence of the flow amplitude $A$ on the existence of the previously described symmetric and asymmetric steady solutions. We consider both positive and negative values of $A$, which correspond to flows opposing and aiding flame propagation, respectively. To better compare results across varying values of $A$, we use the effective propagation speed $\widetilde{U} \equiv U+2A/3$ introduced earlier. Illustrative results are shown in Fig.~\ref{fig:VDAn0CCurves} for selected values of $A$, increasing from top to bottom. The case $A=0$ from Fig.~\ref{fig:VDA0CCurves} is included in the middle sub-figure to  highlight the impact of increasing $A$. In doing so, some portions of the solid curves corresponding to symmetric double-headed unstable flames have been omitted to reduce clutter. Such double-headed flames are also excluded for $A=-2$ and $A=2$. 

\begin{figure}[h!]
\centering
\includegraphics[width=192pt]{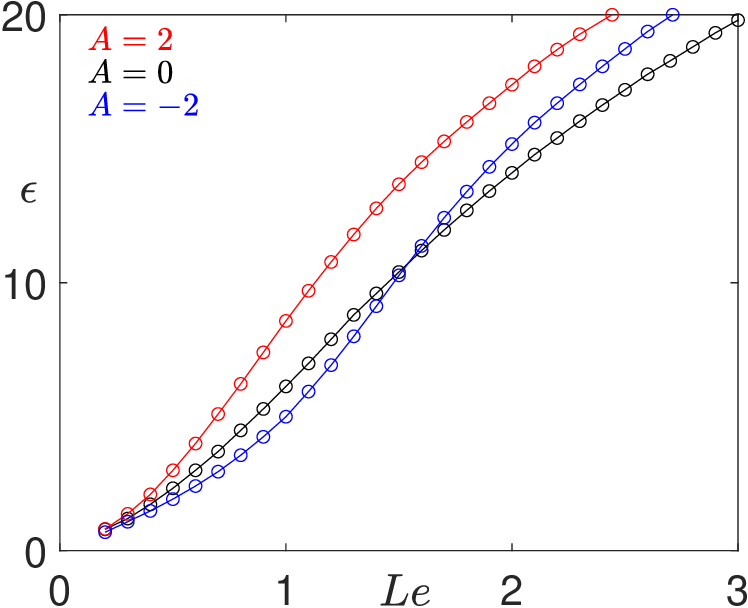}
\caption{\footnotesize The quenching distance $\epsilon$ versus the Lewis number $\Lew$ for selected values of $A$.}
\label{fig:VDLeCriteps}
\end{figure}
\begin{figure}[h!]
\centering
\includegraphics[width=192pt]{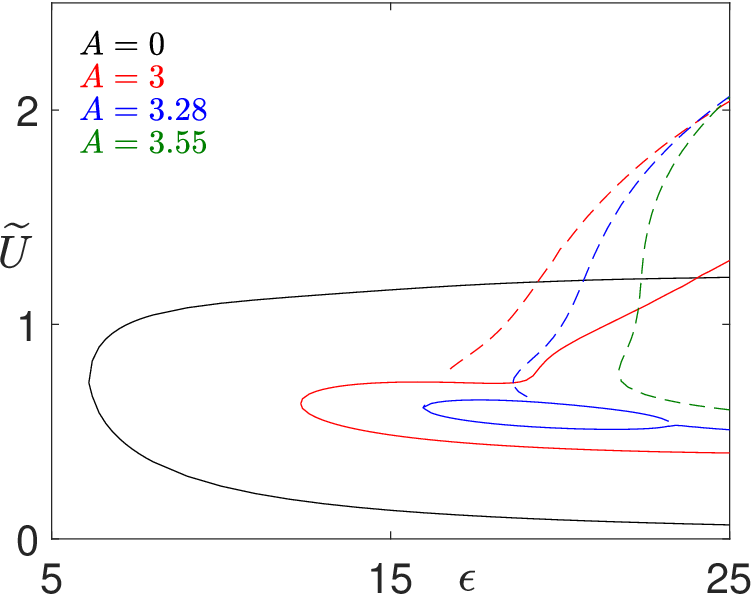}
\caption{\footnotesize  The effective propagation speed $\widetilde{U}$ versus $\epsilon$
for $\Lew=1$ and selected values of $A$. Dashed lines indicate that the solution is asymmetric.}
\label{fig:VDCritasym}
\end{figure}

Inspection of Fig.~\ref{fig:VDAn0CCurves} reveals that an increase in $A$ promotes the appearance of asymmetric solutions. For example, when $A=2$, asymmetric solutions, represented by dashed lines, are observed for $\Lew=0.3$, $0.7$, and $1$. In contrast, for $A=0$, they appear only for $\Lew=0.3$ and $0.7$, while for $A=-2$, they are found only for $\Lew=0.3$. Furthermore, the figure indicates that the quenching distance increases as $A$ is increased, except for the case $Le=2$ considered in this figure (or  more generally, except  for  sufficiently large values of $\Lew$ as indicated in the next figure).
  
The trend  identified for the quenching distance is confirmed in Fig.~\ref{fig:VDLeCriteps}, where the quenching distance is determined versus the Lewis number for three selected values of $A$. The figure confirms that flows opposing flame propagation ($A>0$),  such as for the case $A=2$, lead to an increase in the quenching distance compared to the no-flow case ($A=0$). The extent of this increase is enhanced for larger values of the Lewis number. For flows aiding flame propagation ($A<0$), such as for $A=-2$, the quenching distance is instead decreased compared to the $A=0$ case, provided that $\Lew$ is not too large, specifically $\Lew\lesssim1.55$ when $A=-2$.

An interesting original observation is worth reporting for larger positive values of $A$ in reference to Fig.~\ref{fig:VDCritasym}: asymmetric solutions represented by dashed lines now exist for values of $\epsilon$ where no symmetric solutions exist. That is, the quenching distance is determined by the asymmetric solution, as seen for $A=3.55$, rather than by the symmetric solutions. This
point is further explored in the next section.

%\newpage

\subsection{The quenching distance and flame symmetry \label{sec:ResultsQD}} 

\begin{figure}[h!]
\centering
\includegraphics[width=192pt]{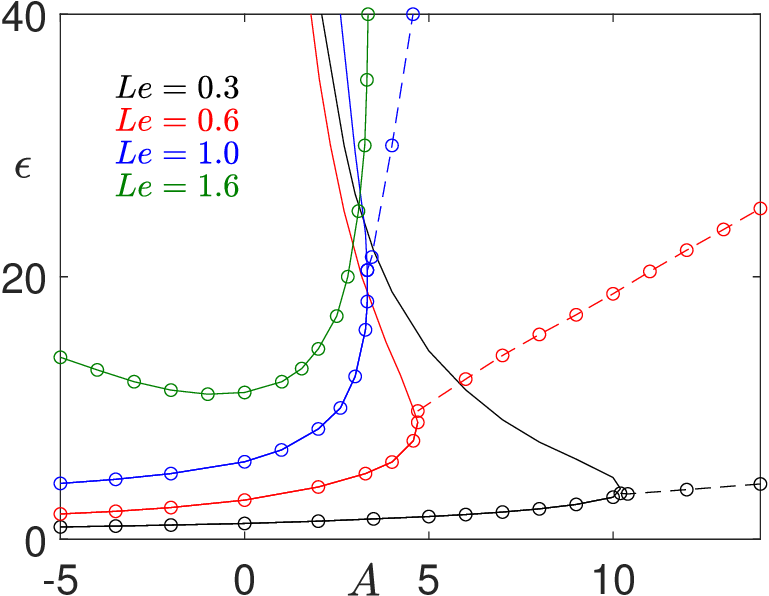}
\caption{\footnotesize The quenching distance $\epsilon$ versus the flow amplitude $A$ for selected values of $\Lew$: curves marked by small circles.
The solid curves delimit the existence domain of symmetric solutions. The dashed curves are determined by asymmetric solutions.}
\label{fig:VDACriteps}
\end{figure}

This section is dedicated to the determination of the quenching distance, with special emphasis on whether it is determined by the symmetric or asymmetric solutions. A summary of the calculations is presented in Fig.~\ref{fig:VDACriteps}, where the quenching distance is plotted as a function of the flow amplitude $A$ for selected values of the Lewis number $\Lew$. Each quenching curve, associated with a specific $\Lew$, is marked by small circles.
Of course, burning steady solutions exist only above the quenching curves.  The solid segments of these curves are determined by symmetric solutions, while the dashed segments are determined by asymmetric solutions. It is important to point out that if only symmetric solutions were considered, say by enforcing numerically a symmetry condition with respect to the plane $y=0$, then the quenching curves would be determined by the (inverse-C shaped) solid curves plotted. This would imply that no (symmetric) burning solutions   exist for values of $A$ exceeding a critical value $A_c$, however wide the channel is.
This conclusion is consistent with that of Daou and Matalon~\cite{daou2002influence}, who imposed a similar restrictive symmetry condition in a constant-density, unit-Lewis-number numerical study. For illustration,
we note that the critical values $A_c$ correspond to the turning points in the solid curves for $\Lew=0.3, \, 0.6$ and $1$. Our  results demonstrate  that when asymmetric solutions are allowed, steady flames can exist for values of $A$ significantly larger than $A_c$, provided the Lewis number is not too large.  
Finally, we note that for negative values of $A$, the quenching distance is determined by the symmetric solutions. Smaller quenching distances are achieved as $A$ becomes more negative, except for sufficiently large Lewis numbers, as seen in the case $\Lew=1.6$. 

\begin{figure}[h!]
\centering
\includegraphics[width=192pt]{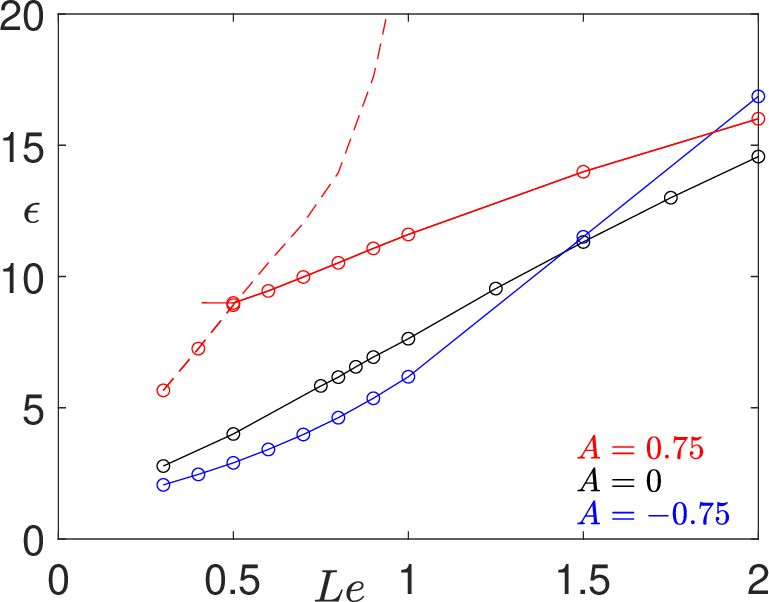}
\caption{\footnotesize  The quenching distance $\epsilon$ versus the Lewis number $\Lew$ for selected values of $A$ within the constant density assumption: curves marked with small circles. Solid lines represent solutions symmetric with respect to $y=0$ and dashed lines asymmetric solutions. The figure is to be compared with  its variable density counterpart, Fig.~\ref{fig:VDLeCriteps}.}
\label{fig:LeCriteps}
\end{figure}
\begin{figure}[h!]
\centering
\includegraphics[width=192pt]{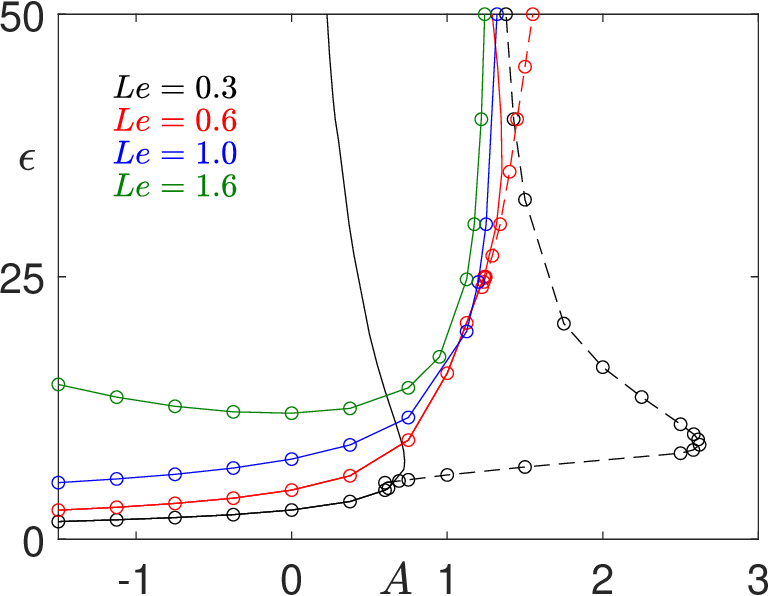}
\caption{\footnotesize The quenching distance $\epsilon$ versus the flow amplitude $A$ for selected values of $\Lew$ within the constant density assumption: curves marked with small circles. 
The solid curves delimit the existence domain of symmetric solutions. The dashed curves are determined by asymmetric solutions. The figure is to be compared with  its variable density counterpart, Fig.~\ref{fig:VDACriteps}.}
\label{fig:ACriteps}  
\end{figure}

\begin{figure}[h!]
\centering
\includegraphics[width=192pt]{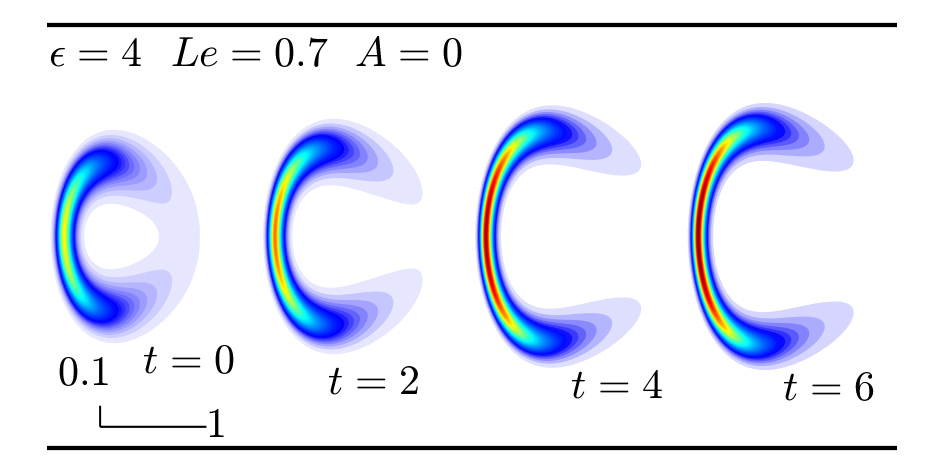}
\caption{\footnotesize  Plots of the reaction rate $\omega$, illustrating the transition from an unstable bottom branch solution (point A in Fig.~\ref{fig:VDLe07A0epsUDiag}) to a stable top branch solution (point B in Fig.~\ref{fig:VDLe07A0epsUDiag}).}
\label{fig:Le07eps4A0wTD}
\end{figure}
\begin{figure*}[h!]
\centering
\vspace{-0.4 in}
\includegraphics[width=408pt]{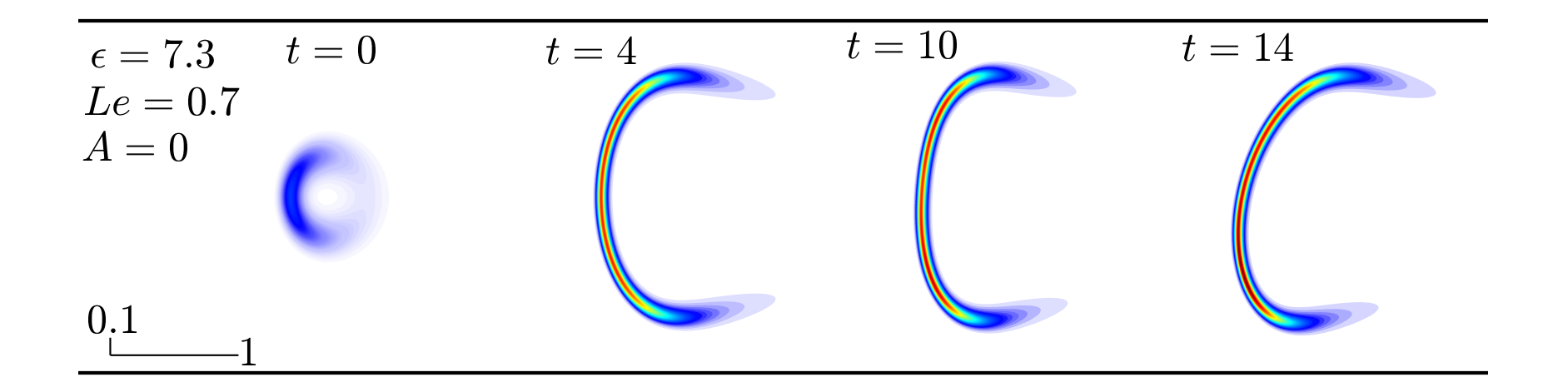}
\vspace{10 pt}
\caption{\footnotesize Plots of the reaction rate $\omega$, illustrating the transition from an unstable bottom branch flame (point C in Fig.~\ref{fig:VDLe07A0epsUDiag}) to large symmetric quasi-stable flame (point D in Fig.~\ref{fig:VDLe07A0epsUDiag}) to a stable asymmetric flame (point E in Fig.~\ref{fig:VDLe07A0epsUDiag}).}
\label{fig:Le07eps73A0wTD}
\end{figure*}
\begin{figure}[h!]
\centering
\includegraphics[width=192pt]{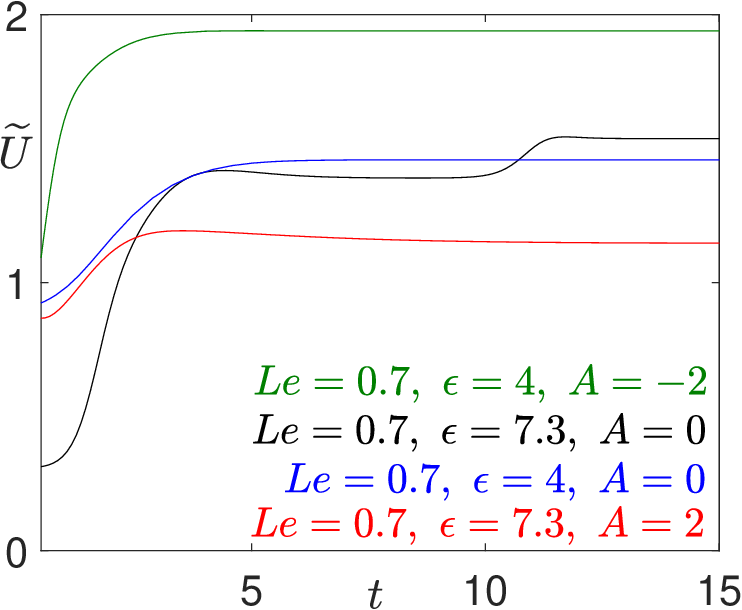}
\vspace*{0.1mm}
\caption{\footnotesize Effective propagation speed $\widetilde{U}$ versus time for selected combinations of $\Lew$, $\epsilon$ and $A$. The green, blue and red curves correspond to direct transitions from a small symmetric to large symmetric flames such as in the case of Fig.~\ref{fig:Le07eps4A0wTD}. The black curve depicts  a transition through a quasi-steady state and corresponds to the case of Fig.~\ref{fig:Le07eps73A0wTD}.}
\label{fig:VDLe07eps73A0Wt}
\end{figure}

\subsection{Quenching distance within the constant density approximation
\label{sec:ResultsConstDen}}

In this section, we briefly examine the influence of the constant-density approximation, an assumption which is often used in theoretical and numerical studies such as \cite{daou2002influence}, on the variable density results presented above. To this end,  our model is reduced into two governing equations for 
$\theta$ and $y_F$ by adopting the diffusive-thermal approximation, and then solved  numerically with a prescribed velocity field. The results are summarised in  Fig.~\ref{fig:LeCriteps} and Fig.~\ref{fig:ACriteps} which are to be compared with Fig.~\ref{fig:VDLeCriteps} and Fig.~\ref{fig:VDACriteps}, respectively. The comparison indicates that the results exhibit qualitatively similar trends, particularly concerning the effects of the flow amplitude and the Lewis number on the quenching distance and the symmetry of steady flames. However, quantitative differences are observed:

(a) The results in the constant density case show heightened sensitivity to the flow amplitude, as evidenced by the smaller values of $A$ featured in Fig.~\ref{fig:LeCriteps} and Fig.~\ref{fig:ACriteps} compared to those in Fig.~\ref{fig:VDLeCriteps} and Fig.~\ref{fig:VDACriteps};  

(b) Asymmetric solutions play a more significant role in determining the quenching curve in the variable density case than in the constant density one. For instance, the quenching curve for $\Lew=1$  is partially determined by asymmetric solutions (dashed segment) in the variable density case but is entirely determined by symmetric solutions (solid line) when the density is assumed constant.

\subsection{Time dependent simulations and flame stability \label{sec:ResultsStab}} 

In this subsection, time-dependent variable-density numerical simulations are presented to examine the stability of the steady solutions described above. This is carried out for several selected cases. The computations are initiated from conditions corresponding to the steady solutions upon which small   random disturbances are superimposed (of relative amplitude $10^{-4}$).

To illustrate the stability of the steady solutions, we begin by 
examining  the case pertaining to point A on the bottom branch in Fig.~\ref{fig:VDLe07A0epsUDiag} corresponding to  $\Lew=0.7$,  $A=0$ and $\epsilon=4$.  Time-dependent simulations initiated from this solution (weakly perturbed by small random perturbations) are carried out, resulting in the instantaneous $\omega$-fields  plotted in Fig.~\ref{fig:Le07eps4A0wTD}. The figure clearly shows that the initial flame presented by point A is unstable, as it evolves in time into a larger flame which persists  indicating that it is stable. We have checked that this larger flame corresponds in fact to the solution presented by point B on the top branch of Fig.~\ref{fig:VDLe07A0epsUDiag}.

We turn now to  Fig.~\ref{fig:Le07eps73A0wTD} which is similar 
to the previous figure, with the initial condition corresponding now to point C on the bottom branch of Fig.~\ref{fig:VDLe07A0epsUDiag} pertaining to $\Lew=0.7,\ A=0$ and $\epsilon=7.3$.  The figure confirms that the point-C steady flame is unstable evolving ultimately to an asymmetric stable flame. The latter is found to  coincide with the flame presented by point E in Fig.~\ref{fig:VDLe07A0epsUDiag}. Interestingly, we observe that the evolution from the initial unstable (small symmetric) flame (point C) to the final stable asymmetric flame (point E) involves a transition through a quasi-stable flame, persisting for a significant time.  This quasi-stable flame is
found to be nearly identical to the large symmetric flame presented by point D in Fig.~\ref{fig:VDLe07A0epsUDiag}. This transition from unstable to stable states through a quasi-steady one is  further confirmed by the black curve in   Fig.~\ref{fig:VDLe07eps73A0Wt}; the curve represents $\widetilde{U}$ as a function of time with its intermediate plateau for $4<t<10$ being associated with the quasi-steady state. In this figure, we have also plotted three other curves for selected values of $\Lew$, $\epsilon$ and $A$, which  illustrate the transition from small symmetric unstable flames to large symmetric ones.
In particular, the blue curve shows the evolution of $\widetilde{U}$ versus time for the case of Fig.~\ref{fig:Le07eps4A0wTD}.

  These findings
have been further validated for constant-density flames through additional computations, though these results are not reported in this work.

\section{Conclusions\label{sec:conclusions}} 

In this study, we have investigated the impact of the Lewis number and flow amplitude on the existence, structure and stability of flames in channels of varying width, with main focus on the determination of the quenching distance.  It is found that smaller Lewis numbers and flows aiding flame propagation ($A<0$) lead to smaller quenching distances, enabling flame propagation in narrower channels. Conversely, opposing flows ($A > 0$) and larger Lewis numbers  lead to larger quenching distances.

Furthermore, it is found that the existence of asymmetric flames is favoured by a reduction in the Lewis number or an increase in the amplitude of opposing flows. This leads to the emergence of asymmetric flames even at unit Lewis numbers, for sufficiently large positive values of $A$.

The investigation also leads to a pertinent observation: 
While quenching distances are typically determined by the symmetric solutions (e.g., the solid-line curves in Fig.~\ref{fig:VDAn0CCurves}), for sufficiently large positive values of $A$, the quenching distance is determined by asymmetric solutions (e.g., the dashed lines in Fig.~\ref{fig:VDACriteps}).  Finally, the following main conclusions regarding the stability of the steady solutions can be made: (a)  the symmetric solutions are unstable in the range of parameter values for which the asymmetric solutions exist, (b) outside this range, the stronger burning symmetric solutions (on the upper branches of the  solid curves in Fig.~\ref{fig:VDAn0CCurves} e.g.) are stable, (c) in all cases, the weaker burning solutions are unstable.

\section*{Declaration of competing interest} 

The authors declare that they have no known competing financial interests or personal relationships that could have appeared to influence the work reported in this paper.

\section*{Acknowledgments} 

This work was supported by the UK EPSRC through grant EP/V004840/1 and  Grant No. APP39756. The authors would like to express their gratitude to Prof.~P.D.~Ronney for suggesting the core topic of this research.

\bibliographystyle{elsarticle-num}

\bibliography{elsarticle-template}

\end{document}